\documentclass[10pt,a4paper]{article}
\usepackage{graphicx}
\usepackage{dcolumn}
\usepackage{authblk}
\usepackage[usenames]{color}
\usepackage{bm}
\usepackage{amsmath}
\usepackage{amssymb}
\usepackage[normalem]{ulem}
\usepackage{lineno}
\usepackage[margin=1in,footskip=0.25in]{geometry}

\newcommand{\bti}{Bi$_2$Te$_2$I$_2$}

\newcommand{\bise}{Bi$_2$Se$_3$}
\newcommand{\bite}{Bi$_2$Te$_3$}
\newcommand{\sbte}{Sb$_2$Te$_3$}

\begin{document}

\title{Quantum spin Hall insulators in centrosymmetric thin films composed from topologically trivial BiTeI trilayers}

\author[1,2,3]{I.~A.~Nechaev}
\author[2,3,4,5]{S.~V.~Eremeev}
\author[5,6,7]{E.~E.~Krasovskii}
\author[5,6]{P.~M.~Echenique}
\author[1,2,3,5,6]{E.~V.~Chulkov}

\affil[1]{Centro de F\'{i}sica de Materiales CFM - MPC and Centro Mixto CSIC-UPV/EHU, 20018 San Sebasti\'{a}n/Donostia, Spain}
\affil[2]{Tomsk State University, 634050 Tomsk, Russia}
\affil[3]{Saint Petersburg State University, 198504 Saint Petersburg,  Russia}
\affil[4]{Institute of Strength Physics and Materials Science, 634055 Tomsk, Russia}
\affil[5]{Donostia International Physics Center, 20018 San Sebasti\'{a}n/Donostia, Spain}
\affil[6]{Departamento de F\'{i}sica de Materiales UPV/EHU, Facultad de Ciencias Qu\'{i}micas, UPV/EHU, Apdo. 1072, 20080 San Sebasti\'{a}n/Donostia, Spain}
\affil[7]{IKERBASQUE, Basque Foundation for Science, 48013 Bilbao, Spain}

\maketitle

\begin{abstract}
{\bf
The quantum spin Hall insulators predicted ten years ago and now experimentally observed are instrumental for a breakthrough in nanoelectronics due to non-dissipative spin-polarized electron transport through their edges. For this transport to persist  at normal conditions, the insulators should possess a sufficiently large band gap in a stable topological phase. Here, we theoretically show that quantum spin Hall insulators can be realized in ultra-thin films constructed from a trivial band insulator with strong spin-orbit coupling. The thinnest film with an inverted gap large enough for practical applications is a centrosymmetric sextuple layer built out of two inversely stacked non-centrosymmetric BiTeI trilayers. This nontrivial sextuple layer turns out to be the structure element of an artificially designed strong three-dimensional topological insulator \bti. We reveal general principles of how a topological insulator can be composed from the structure elements of the BiTe\textit{X} family (\textit{X}=I, Br, Cl), which opens new perspectives towards engineering of topological phases. 
}
\end{abstract}

Two-dimensional (2D) topological insulators (TIs)---a new electronic phase also referred to as a quantum spin Hall (QSH) insulator---are characterized by an absolute band gap induced by spin-orbit coupling (SOC) and helical gapless edge states inside the gap \cite{Qi_RMP_2011}. These states protected by time-reversal symmetry provide perfectly conducting spin-filtered channels, meeting the demands of low-power nanoelectronics and spintronics.
The existence of such states as the fingerprint of a topologically non-trivial 2D insulator was first predicted in Refs.~\cite{Kane_Mele_PRL_1, Kane_Mele_PRL_2}. It was also suggested that the QSH effect can be observed in graphene, where SOC opens a gap at the two inequivalent Dirac points. This gap in graphene appears to be too small for practical use, so heavy-elements based analogs of graphene must be sought. Actually,  the 2D materials with low-buckled honeycomb-lattice structures \cite{Cahangirov_PRL_2009, Lalmi_APL_2010, Liu_PRL_2011, Liu_PRB_2011, Ezawa_NJP_2012, Davila_NJP_2014, Zhu_NatMat_2015}---silicene, germanene, and stanene---possess a significantly larger SOC-induced gap at the Dirac points (up to $\sim0.1$ eV in stanene), and the spin-polarized edge channels could be detected at easily accessible temperatures. However, the QSH effect in such systems has not been experimentally observed so far.

Further effective enhancement of the SOC to make the gap larger can be realized by chemical functionalization of the above 2D materials \cite{Ma_GeI_2012, Xu_PRL_2013}. Such a functionalization substantially enlarges the gap, in fact ``destroying'' the Dirac cones, and it may lead to a SOC-induced band inversion at the time reversal invariant momentum (TRIM) (normally at \textbf{k}=0) with an absolute gap of several hundred meV at this momentum. If the band inversion occurs, the resulting 2D system is a 2D TI that should support the QSH effect \cite{Xu_PRL_2013} similar to the inverted HgTe quantum wells predicted to be QSH insulators in Ref.~\cite{Bernevig_Science_2006}. It is important that this prediction has found experimental confirmation: In the inverted HgTe/CdTe and InAs/GaSb quantum wells \cite{Konig, Knez,Koenig_JPSJ_2008} the QSH effect was observed despite the very small gaps in these quantum wells, less than 10 meV. It has spurred a rising tide of theoretical propositions of different 2D TIs with honeycomb- or square-lattice structures and a large inverted gap enabling room-temperature operating \cite{Ren_ROPP_2016}.

\textit{Ab initio} approaches to electronic structure, especially those based on the density functional theory (DFT), have become a powerful tool to search for new materials with unique properties. At the same time, the effective models that proved indispensable in predicting the QSH effect in graphene-like systems and quantum wells are currently widely used to analyze the effect of strain, quantum confinement, and external fields in 2D TIs, i.e., to solve the problems that presently are not accessible with \textit{ab initio} methods. Thus, to efficiently model the nanoelectronics and spintronics devices, the microscopic methodology must be bridged with the effective Hamiltonian approach based on symmetry considerations and on the \textbf{k$\cdot$p} perturbation theory.

With a few exceptions, none of the theoretically proposed 2D materials has been hitherto fabricated \cite{Ren_ROPP_2016}. Thus, the intensive search of robust and easily fabricated materials remains to be actual. In particular, it was suggested that 2D TIs can be produced from a thin film of
layered 3D (three-dimensional) TIs of the \bise\ family, where the hybridization between the opposite surfaces of the film opens a gap at the Dirac point (DP). Depending on film thickness, the 2D system may ``oscillate'' between band insulator and QSH insulator as was predicted by the 4-band effective \textbf{k$\cdot$p} model (see Refs. \cite{Lu_PRB_2010,Liu}). The thinnest known topologically nontrivial film consists of at least two structural elements---quintuple or septuple layers \cite{Bihlmayer_Book_TI}.

Besides the studies on the thin films of 3D TIs, recently it was heuristically suggested that a 3D TI can be constructed artificially via stacking 2D bilayers that are topologically trivial \cite{DasBalatsky}. It encourages our search for 2D TIs built out of trivial band-insulator constituents. These
constituents should have strong spin-orbit coupling (SOC), and prospective candidates are bismuth tellurohalides BiTe\textit{X} with \textit{X}=I, Br, and Cl, among which the polar semiconductor BiTeI demonstrates the strongest spin-orbit coupling providing the biggest known Rashba spin-splitting of bulk and surface states \cite{Ishizaka,Eremeev_PRL2012}. The structure element of BiTeI is a trilayer (TL) with the I-Bi-Te stacking. A single TL that possesses the Rashba spin-split band structure \cite{Eremeev_SRep_TI-BTX} can be grown epitaxially on a suitable substrate or be easily exfoliated from the bulk BiTeI, where the adjacent TLs couple through a weak van-der-Waals (vdW) interaction. The samples of BiTeI always contain a large number of randomly distributed bulk stacking faults, which leads to a mixture of terminations at the surface, as experimentally observed in  Refs.~\cite{Landolt,Fiedler,Tournier_Colletta,Crepaldi,Butler_NatComm,Fiedler_2}. This implies that adjacent TLs may have different sequence order along the hexagonal \textbf{z} axis.

Here, based on DFT calculations we demonstrate that a centrosymmetric sextuple layer (SL) constructed from two BiTeI TLs with facing Te-layer sides and a typical vdW spacing is a 2D TI with the gap of 70 meV at $\bar{\Gamma}$. The vdW interaction between these TLs is crucial to realize such a QSH insulator phase: The SL becomes topologically trivial with increasing the vdW spacing by 5\% only. We consider the \textit{nontrivial } SL as a structure element, a repetition of which along the \textbf{z} axis results in thin films that are found to ``oscillate'' between trivial and nontrivial phases with the number of SLs. The corresponding bulk system composed of SLs turns out to be a \textit{strong} 3D TI (hereafter referred to as \bti). It is energetically unfavourable by only  0.5~meV compared with the non-centrosymmetric BiTeI. This makes it plausible to suppose that crystals of BiTeI grown by the Bridgman method already contain the desired SLs, and that the alternative stacking can be experimentally observed and controllably manufactured. To describe the low-energy properties of \bti\, and its films, we derive four-band \textbf{k$\cdot$p} Hamiltonians from the \textit{ab initio} wave functions. They are similar to the Hamiltonians constructed for \bise-family 3D TIs and their thin films \cite{Zhang_Nat_Phys_2009, Lu_PRB_2010, Zhang_PRB_2012}. For a more accurate description of the SL, we derive an eight-band Hamiltonian that involves Rashba-split valence and conduction bands of the stand-alone TLs. We thus demonstrate that due to the bonding-antibonding splitting the inversion occurs between one of the Te-related valence bands and one of the conduction bands formed by Bi orbitals. The proposed materials illustrate the effectiveness of the new way to design 2D TIs from trivial band insulators with giant-Rashba spit bands for room-temperature operating.

\section*{Results}

\begin{figure}[t]
  \centering
  \includegraphics[width=\columnwidth]{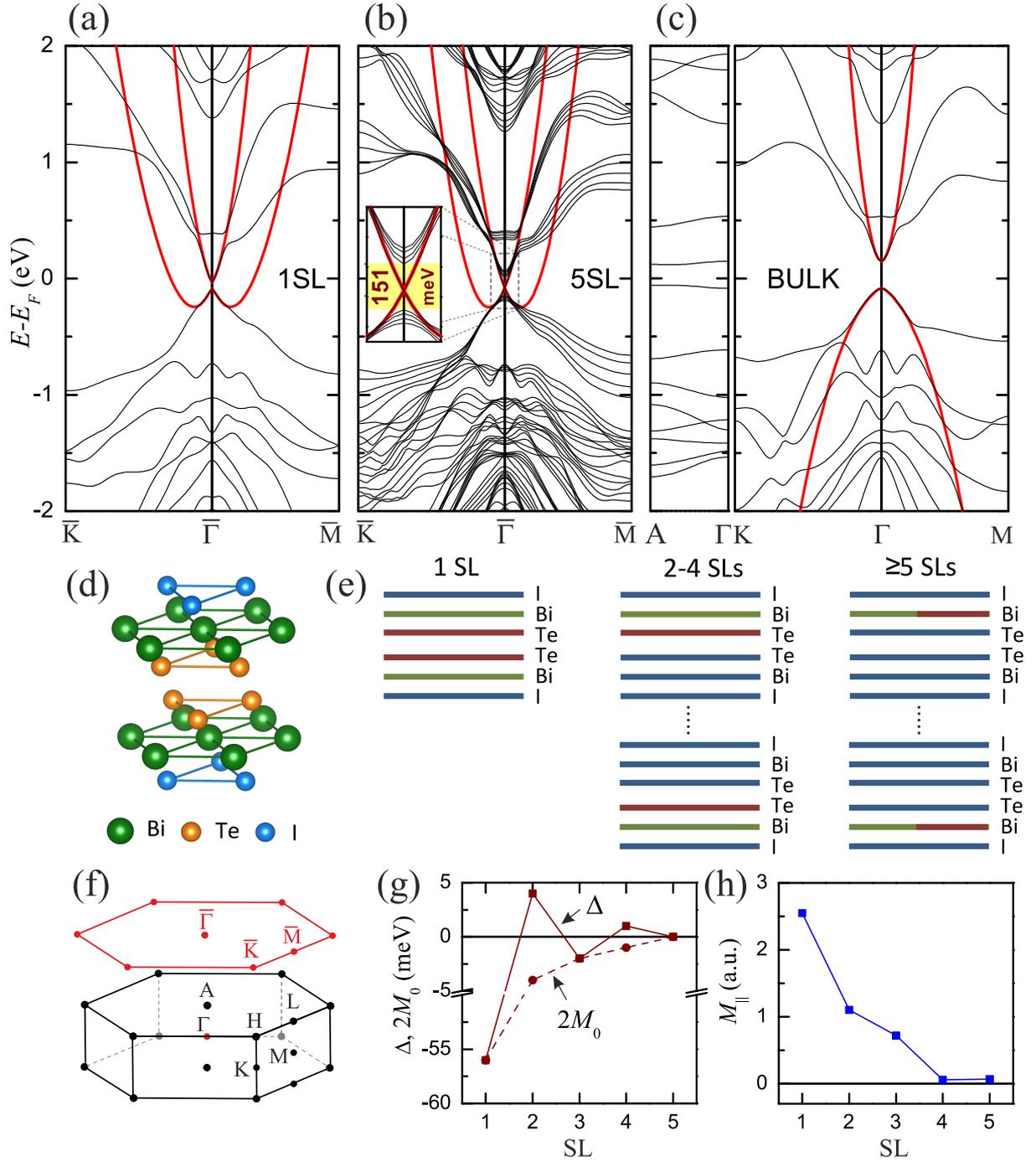}
  \caption{\textbf{\textit{Ab initio} calculations for \bti.} LDA band structure (black curves) of 1SL (a) and 5SL (b) films as well as the bulk crystal (c) of \bti. Red curves correspond to the eigenvalues $E(\mathrm{\mathbf{k}})$ of the model four-band Hamiltonian. Atomic layers maximally contributing to the valence and the conduction band at the center of the 2D Brillouin zone in the case of the \bti\, films of different thickness are indicated in graph (e) by green and red bars, respectively. The dependence of the model Hamiltonian parameters on the number of SLs is presented in graphs (g) and (h). The atomic structure of the bulk \bti\, (unit cell) and its 1SL film (d) with the corresponding 3D and 2D Brillouin zone (f) are also shown.}
  \label{bulk_5SL_params}
\end{figure}

Figures~\ref{bulk_5SL_params}(a)-\ref{bulk_5SL_params}(c) show the band structure of 1 and 5 SLs and the bulk crystal of \bti\, obtained with the extended linearized augmented plane wave (ELAPW) method \cite{Krasovskii_PRB_1997} within the local density approximation (LDA) for the exchange-correlation functional and with the use of the full potential scheme of Ref.~\cite{Krasovskii_PRB_1999}. (Details on the equilibrium bulk atomic structure, the bulk-truncated slab geometry of the related thin films, and the calculations performed can be found in Supplementary Note 1.)

\begin{figure}[t]
  \centering
  \includegraphics[width=\columnwidth]{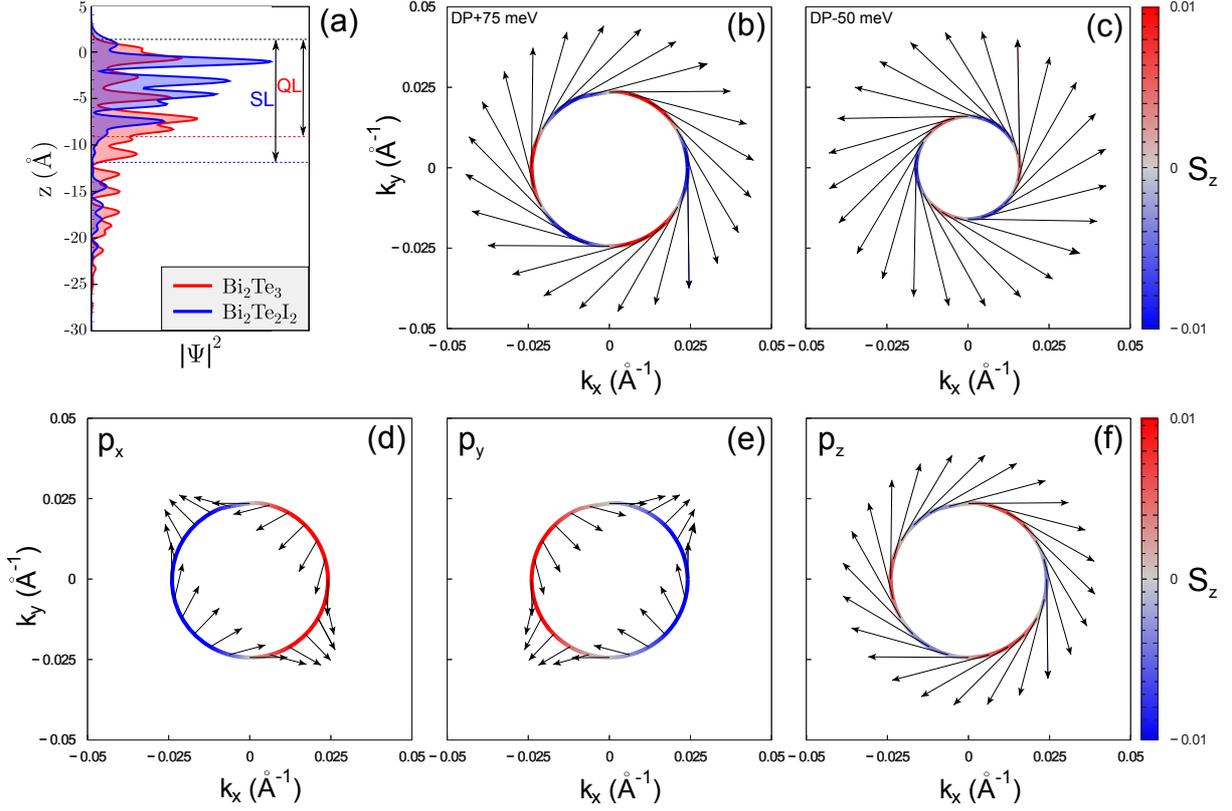}
  \caption{\textbf{Spin-orbital texture of the surface state.} (a) Spatial localization of the topological surface state as function
  of $z$, where $z=0$ corresponds to the surface plane for \bti\ and \bite.   Spin-resolved constant energy contours taken at
  75 meV above (b) and at 50 meV below (c) the Dirac point (DP). Black arrows adjacent to the contours denote the
   in-plane spin component $S_\|$; the out-of-plane spin component $S_z$ is indicated by the color in contour, with
   red and blue corresponding to positive and negative values, respectively. Projections of the total spin for upper Dirac cone,
   shown in panel (b), on the $p_x$ (d), $p_y$ (e), and $p_z$ (f) orbitals.}
 \label{5SL_spin}
\end{figure}

The 1SL film is constructed from two BiTeI trilayers with facing Te-layer sides, Fig.~\ref{bulk_5SL_params}(d). It is noteworthy that the band structure of this film with the gap of 56 meV (70 meV in the relaxed geometry, see Supplementary Fig.~1) differs substantially from that of its constituents (cf. Supplementary Fig.~2): there is no trace of Rashba-type split bands. The band structure of the 5SL film exhibits a gapless Dirac state residing in the band gap of 151~meV, see Fig.~\ref{bulk_5SL_params}(b) and Supplementary Fig.~3. This is a signature of the topological character of the respective bulk band structure (Fig.~\ref{bulk_5SL_params}(c)), which has an inverted gap of 234~meV at $\Gamma$ and the fundamental gap of 169~meV in the $\Gamma-$A line close to the A point. As seen in Fig.~\ref{5SL_spin}(a), the Dirac surface state almost completely resides within the outer SL. Moreover, this state is localized stronger than the Dirac state of TIs like \bite, since 70\% of its weight falls in the outermost half of the SL, i.e., in  the surface BiTeI TL (see also Fig.~\ref{bulk_5SL_params}(e)).

The spin texture of the Dirac state is illustrated in Figures~\ref{5SL_spin}(b) and \ref{5SL_spin}(c), which show spin-resolved constant energy contours for the lower and upper cones of the Dirac surface state. Apart from the in-plane polarization -- clockwise above the DP and counterclockwise below it,
both contours also have an out-of-plane spin component, which is an intrinsic feature of the hexagonal surface. However, in this case $S_z$ is extremely small and varies in the range of $\approx \pm 0.01$.

As has been shown for the Dirac state in \bise\ both experimentally and theoretically \cite{Zhang_PRL_2013,Jozwiak_NatPhys_2013,Cao_NatPhys_2013,Zhu_PRL_2014,Xie_NatComm_2014}, the spin textures of $p_x$, $p_y$, and $p_z$ orbitals are remarkably different, which leads to the dependence of the spin polarization of photoelectrons on the polarization of light. The spin texture provided by $p_z$ orbitals has clockwise (counterclockwise) chirality for the upper (lower) cone, while the projections of the total spin on $p_x$ and $p_y$ orbitals are not chiral, and their spins are opposite to each other. Similar spin-orbital texture we find in the \bti, see Figures~\ref{5SL_spin}(d)--\ref{5SL_spin}(f) for the upper Dirac cone (for the lower Dirac cone, the coupling of spin and orbital textures is opposite, not shown). As can be seen, the spin orientations for $p_x$ and $p_y$ projections are antiparallel at each $k_\|$ point, whereas the spin orientation of $p_z$ projection coincides with the total spin.

It is noteworthy that similar spin-orbital texture has been observed for the spin-polarized Rashba state in BiTeI \cite{Maass_NatComm_2016}. For the Te-terminated surface, it was found that the outer Rashba branch demonstrates the same spin orientations for $p_x$, $p_y$, and $p_z$ projections as  those for the upper Dirac cone in \bti, and the inner Rashba branch has opposite texture, i.e., the same as in the lower cone. Because the surface of the  \bti\ slab has iodine termination and its spin-texture is reversed due to the opposite orientation of the \textbf{z} axis, the spin-orbit texture of the upper
(lower) Dirac cone in \bti\ is the same as the texture of the inner (outer) Rashba branch in BiTeI.

To construct a simple effective \textbf{k$\cdot$p} model for the centrosymmetric \bti\, we derive a model Hamiltonian of a desired dimension and accurate up to the second order in \textbf{k} from the LDA spinor wave functions $\Psi_{n\uparrow(\downarrow)}$ of the doubly degenerate bands $E_n$ found at \textbf{k}=0 (see Supplementary Note 2 and Ref.~\cite{NeKras_PRBR_2016} for details. The subscripts $\uparrow$ or $\downarrow$ in $\Psi_{n\uparrow(\downarrow)}$ refer to the $z$-component of the total angular momentum $\mathrm{\mathbf{J}}=\mathrm{\mathbf{L}}+\mathrm{\mathbf{S}}$  in the atomic sphere that has the largest weight in the $n$-th band, see Fig.~\ref{bulk_5SL_params}(e). The Hamiltonian is constructed in terms of the matrix elements \cite{Krasovskii_PRB_2014} $\bm{\pi}_{n\uparrow(\downarrow)  m\uparrow(\downarrow)}=\langle\Psi_{n\uparrow(\downarrow)}| \bm{\pi}|\Psi_{m\uparrow(\downarrow)}\rangle$ of the velocity operator
$\bm{\pi}=-i\hbar\mathrm{\bm{\nabla}}+\hbar\left[\bm{\sigma} \times \mathrm{\bm{\nabla}} V\right]/4m_0c^2$, where $n$ and $m$ run over the relativistic bands (from semi-core levels up to high-lying unoccupied bands). Here, $\bm{\sigma}$ is the vector of  the Pauli matrices, and $V(\mathrm{\mathbf{r}})$ is the crystal potential.

For the bulk \bti, in the basis of the two valence bands $\Psi^{\mathrm{bulk}}_{v\uparrow (\downarrow)}$ and two conduction bands $\Psi^{\mathrm{bulk}}_{c\uparrow (\downarrow)}$, our \textit{ab initio} four-band Hamiltonian reads:
\begin{eqnarray}\label{Hkp_bulk}
H^{\mathrm{bulk}}_{\mathrm{\mathbf{kp}}}&=&C\tau_0\sigma_0+M\tau_z\sigma_0 \\
&-&V_{\|}\tau_x(\sigma_xk_y-\sigma_yk_x)-V_z\tau_y\sigma_0k_z, \nonumber
\end{eqnarray}
where $C=C_0+C_zk_z^2+C_{\|}k_{\|}^2$, $M=M_0+M_zk_z^2+M_{\|}k_{\|}^2$, $k_{\|}^2=k_{x}^2+k_{y}^2$, and the direct matrix product of the Pauli matrices $\bm{\tau}$ and $\bm{\sigma}$ is implied (the explicit matrix form of $H^{\mathrm{bulk}}_{\mathrm{\mathbf{kp}}}$ is presented in Supplementary Note 3). Note that this Hamiltonian is the same (to within a unitary transformation) as that constructed for \bise\, in Ref.~\cite{Zhang_Nat_Phys_2009} within the theory of invariants. The matrices $\bm{\tau}$ and $\bm{\sigma}$ in Eq.~(\ref{Hkp_bulk}) have different meaning: $\bm{\tau}$ operates in the valence-conduction band space, while $\bm{\sigma}$ refers to the total angular momentum $\mathrm{\mathbf{J}}$.

The parameters in Eq.~(\ref{Hkp_bulk}) obtained within the LDA are: $C_0=0.03$~eV, $C_z=0.13$~a.u., $C_{\|}=4.19$~a.u., $M_0=-0.12$~eV, $M_z=1.35$~a.u., $M_{\|}=5.88$~a.u., $V_{\|}=0.52$~a.u., and $V_z=0.13$~a.u. (we use Rydberg atomic units: $\hbar=2m_0=e^2/2=1$).  Since the basis functions explicitly refer to the valence and conduction bands rather than to atomic orbitals, the parameter $M_0$ that defines the band gap at \textbf{k}=0 is negative and does not change sign upon moving from the topologically non-trivial insulator to the trivial one. The eigenvalues  $E(\mathrm{\mathbf{k}})$ of the Hamiltonian (\ref{Hkp_bulk}) with the above parameters are shown in Fig.~\ref{bulk_5SL_params}(c) by red lines, nicely reproducing the LDA curves over a quite large $\mathrm{\mathbf{k}}$-region and providing an absolute gap in the \textbf{k$\cdot$p} spectrum.
Moreover, these parameters reflect the band inversion and meet the condition of the existence of topological surface states (see, e.g., Ref.~\cite{Shan_NJP_2010}) in accord with the $\mathbb{Z}_2$ topological invariant $\nu_{\rm 3D}=1$ obtained from the parities of the bulk LDA wave functions at the TRIM points \cite{Fu_PRB_2007}. Actually, the diagonal dispersion term $M_{z(\|)}$ is positive, and it is larger than the electron-hole asymmetry: $|C_{z(\|)}|<M_{z(\|)}$.

For the \bti\, thin films, we derive the Hamiltonian in the basis $\Psi^{\mathrm{slab}}_{v\uparrow}, \Psi^{\mathrm{slab}}_{c\downarrow}, \Psi^{\mathrm{slab}}_{c\uparrow},\Psi^{\mathrm{slab}}_{v\downarrow}$ as
\begin{equation}\label{Hkp_slab}
H^{\mathrm{slab}}_{\mathrm{\mathbf{kp}}}=C\tau_0\sigma_0+M\tau_z\sigma_z-V_{\|}\tau_0(\sigma_xk_y-\sigma_yk_x),
\end{equation}
where $C=C_0+C_{\|}k_{\|}^2$, $M=M_0+M_{\|}k_{\|}^2$, and $\bm{\tau}$ refers now to the two decoupled sets of massive Dirac fermions. The Hamiltonian (\ref{Hkp_slab}) is similar to the one obtained for 3D TI thin films within the effective continuous model based on the substitution
$k_z \to -i{\partial }_z$  in the Hamiltonian of Ref.~\cite{Zhang_Nat_Phys_2009} and on the imposition of the open boundary conditions (see, e.g., Refs.~\cite{Lu_PRB_2010} and \cite{Shan_NJP_2010}). The crucial difference is that in our \textit{ab initio} approach within the same formalism for 3D and 2D systems we obtain the Hamiltonian and its parameters from the original spinor wave functions. We do not \textit{a priori} impose the form of the Hamiltonian based on symmetry arguments and do not resort to the fitting of \textit{ab initio} band dispersion curves or to a solution of 1D Schr\"{o}dinger equations derived by using the above substitution with special boundary conditions.

All the considered \bti\, films are characterized by the velocity $V_{\|}=0.45\pm0.01$ a.u. and the electron-hole asymmetry $C_{\|}=4.15\pm0.10$ a.u., which are weakly sensitive to the number of SLs, where the $\pm$ ranges indicate the variations of $V_{\|}$ and $C_{\|}$ in moving from 1 to 5 SLs. On the contrary, as seen in Figures~\ref{bulk_5SL_params}(g) and \ref{bulk_5SL_params}(h) the parameters $M_0$ and $M_{\|}$ depend strongly on the film thickness, approaching monotonically zero.

In order to explicitly indicate whether a given film is a QSH insulator, in Fig.~\ref{bulk_5SL_params}(g) we also plot the gap parameter $\Delta=2M_0(-1)^{1+\nu_{2D}}$  with $\nu_{2D}$ being the $\mathbb{Z}_2$  invariant obtained from the parities of the wave functions at the TRIM points of the 2D Brillouin zone. This parameter is negative for a topologically non-trivial film and positive for a trivial one. As follows from the figure,
$\Delta$ ``oscillates'' with the period of 2 SLs within the examined thickness interval. (The parity of $\Psi^{\mathrm{slab}}_{v\uparrow(\downarrow)}$ and $\Psi^{\mathrm{slab}}_{c\uparrow(\downarrow)}$  is $(+)$ and $(-)$, respectively, for $\Delta<0$, and it is $(-)$ and $(+)$ for $\Delta>0$.) As in 3D TI films \cite{Foerster_PRB_2016}, the thickness dependence of $\Delta$ may be sensitive to the quasi-particle approximation employed, and it may change if many-body corrections beyond DFT are introduced. However, even the simplest quasi-particle method, the $GW$ approximation for the self-energy, is  methodologically challenging and computationally too demanding to study a large series of complex systems. Thus, DFT remains the method of choice, and its good performance for a wide range of TIs justifies the use of the Kohn-Sham band structure as a reasonable starting point.

The diagonalization of the Hamiltonian (\ref{Hkp_slab}) then leads to $E(\mathrm{\mathbf{k}})$  shown by red lines in Figures~\ref{bulk_5SL_params}(a) and \ref{bulk_5SL_params}(b). The absence of the absolute gap in the resulting \textbf{k$\cdot$p} spectrum is the general feature of all the films studied. It is caused by the rather big electron-hole asymmetry $C_{\|}$ compared with the diagonal dispersion parameter $M_{\|}$,  Fig.~\ref{bulk_5SL_params}(h). It should be noted that the conclusion on whether the edge states exist in a TI film is often made based on the signs  and relative values of the parameters $M_0$, $M_{\|}$, and $C_{\|}$. On the contrary, we find that the asymmetry $|C_{\|}|$ is larger than $|M_{\|}|$  everywhere, breaking one of the conditions for the film to be a QSH insulator, see, e.g., Refs.~\cite{Zhou_PRL_2008} and \cite{Lu_PRB_2010}.
Focusing on the behaviour of the diagonal dispersion $M_{\|}$ (as, e.g., in the topology analysis of Ref.~\cite{Koenig_JPSJ_2008}), we note
that it is positive for all the thicknesses, Fig.~\ref{bulk_5SL_params}(h). Along with the negative $M_0$, this should signify an inverted band gap for the respective films. However, it does not correlate with the oscillating $\Delta$, Fig.~\ref{bulk_5SL_params}(g).

\begin{figure}[t]
  \centering
  \includegraphics[width=\columnwidth]{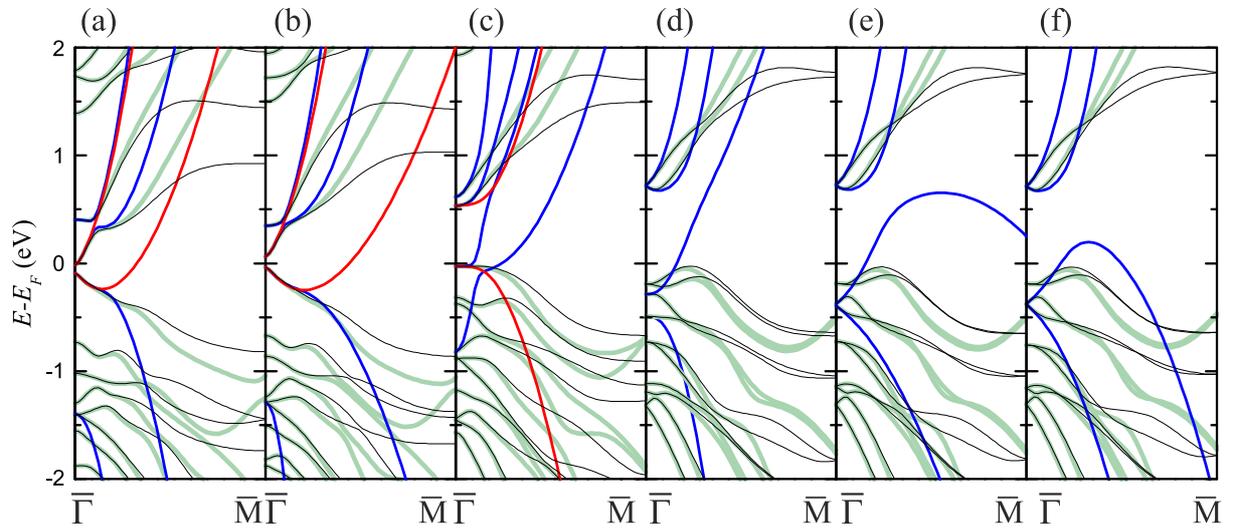}
  \caption{\textbf{Band structure of the 1SL \bti\, film with different van-der-Waals spacing.} The expansion of the spacing is 10\% (b), 40\% (c), 100\% (d), 200\% (e), 1100\% (f) as indicated in percents of the bulk value, (a). The spectra provided by the 4-band (red lines) and 8-band (blue lines) \textbf{k$\cdot$p} models are presented in comparison with the LDA bands (black lines). Light green lines show the results of the 36-band Hamiltonian to demonstrate the convergence with respect to the size of the basis set.}\label{1SL_vs_vdW_bands}
\end{figure}

\begin{figure}[t]
  \centering
  \includegraphics[width=\columnwidth]{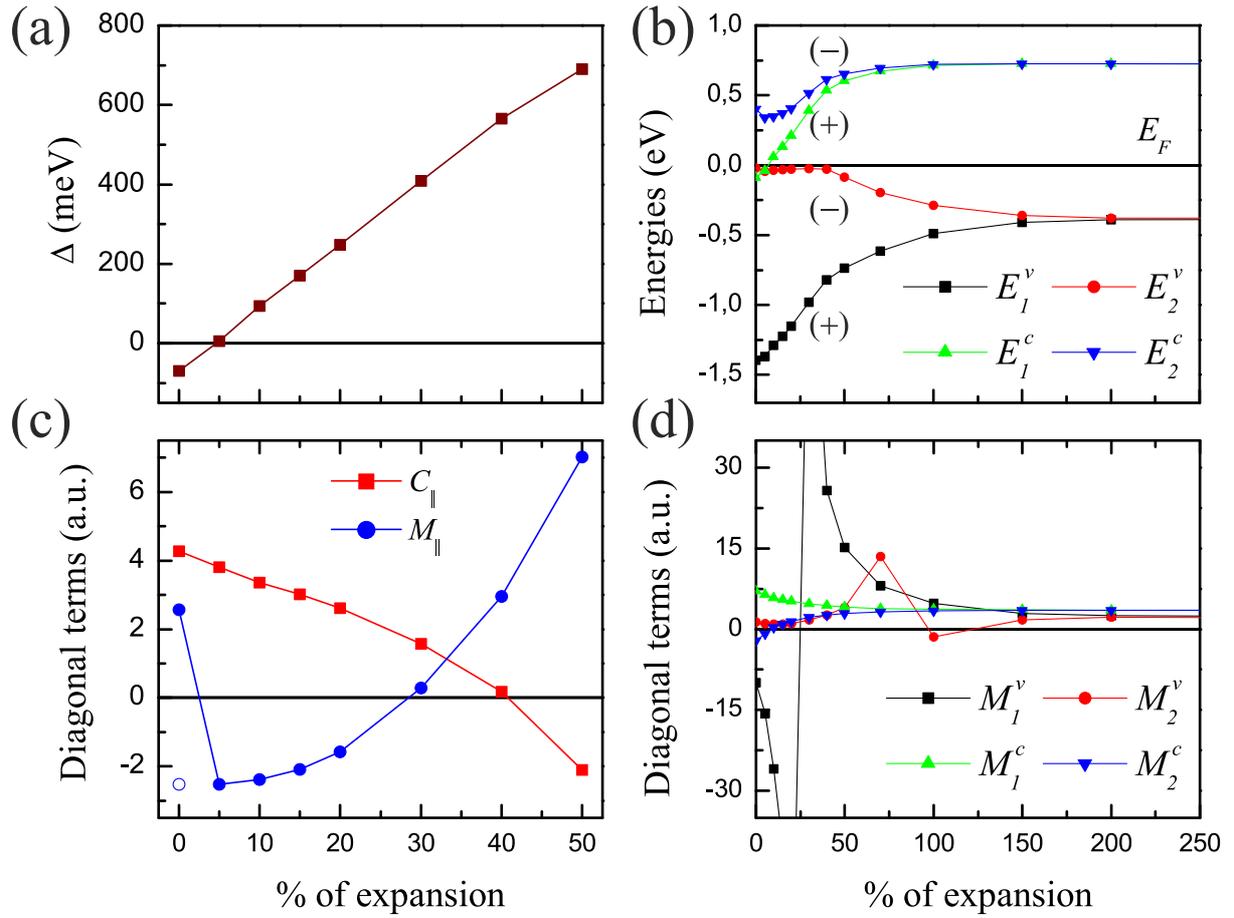}
  \caption{\textbf{The parameters of the \textbf{k$\cdot$p} Hamiltonian for the 1SL \bti\, film.} The parameters of the 4-band (a, c) and 8-band (b, d) Hamiltonian are shown as a function of the expansion of the van-der-Waals spacing, which is given in percents of the bulk value. In graph (b), the parity of the respective LDA wave functions is also
  shown.}\label{1SL_vs_vdW_params}
\end{figure}

Let us now analyze the behaviour of the diagonal dispersion parameter $M_{\|}$ together with the topological invariant under a continuously varying geometry. We choose the 1SL film--the thinnest film, for which the \textbf{k$\cdot$p} prediction of the band inversion does not contradict the actual topological property--and gradually expand the van-der-Waals spacing $d_{\mathrm{vdW}}$.  The evolution of the band structure with increasing
$d_{\mathrm{vdW}}$ is shown in Fig.~\ref{1SL_vs_vdW_bands}. According to the gap parameter $\Delta$, see Fig.~\ref{1SL_vs_vdW_params}(a), a topological phase transition occurs at $d_{\mathrm{vdW}}$ that is just around the mentioned 5\% larger than its bulk value, and the 1SL film becomes topologically trivial. Further expansion leads to a larger band gap at $\bar{\Gamma}$, which is not inverted anymore. It is noteworthy that such a behaviour of $\Delta$ as a function of $d_{\mathrm{vdW}}$ with the topological phase transition around 5\% is stable with respect both to the choice of the approximation to the DFT exchange-correlation functional (LDA, GGA, dispersion corrected GGA) and to the SL geometry (bulk truncated or relaxed). In the limit of very large $d_{\mathrm{vdW}}$, when the BiTeI trilayers composing the 1SL film are too far from each other, the band structure is identical to that of a free-standing BiTeI trilayer (see Supplementary Fig. 2). Similarly, artificial reduction of the spin-orbit interaction strength $\lambda$ relative to its actual value $\lambda_0$ in the equilibrium SL leads to a decrease in the gap, which closes at $\lambda/\lambda_0=0.95$. A further decrease in $\lambda$ causes a widening of the already uninverted gap of the trivial phase. In general, the dependence of the relative gap-width on the spin-orbit interaction strength is almost linear and can be approximated as $\Delta(\lambda)/|\Delta(\lambda_0)|=-20.9(\lambda/\lambda_0)+19.9$.

The 1SL parameters of the 4-band \textbf{k$\cdot$p} Hamiltonian (\ref{Hkp_slab}) strongly depend on $d_{\mathrm{vdW}}$ (the respective eigenvalues $E(\mathrm{\mathbf{k}})$ of this Hamiltonian are shown by red lines in Figures~\ref{1SL_vs_vdW_bands}(a)-\ref{1SL_vs_vdW_bands}(c)). With the $d_{\mathrm{vdW}}$ expansion $\xi$ (given in percents of the bulk value $d^{(0)}_{\mathrm{vdW}}$) up to 50\%, the velocity $V_{\|}$ decreases monotonically from 0.470 a.u. to 0.342~a.u., and the electron-hole asymmetry $C_{\|}$ becomes smaller as well, Fig.~\ref{1SL_vs_vdW_params}(c). At $\xi=33\%$, $C_{\|}$ is already smaller than $M_{\|}$, ensuring an absolute gap in the 4-band \textbf{k$\cdot$p} spectrum, see Fig.~\ref{1SL_vs_vdW_bands}(c). With further increasing $\xi$ it even becomes negative, but it remains $|C_{\|}|<M_{\|}$. A stepwise behaviour of the parameter $M_{\|}$ that changes sign at the small $\xi$ indicates that $M_{\|}$ keeps following the actual $\nu_{\rm 2D}$ and, thus, predicts a gap  without inversion. With increasing $d_{\mathrm{vdW}}$ this parameter again goes through zero around $\xi=30\%$, telling us that the band gap becomes inverted again, and at $\sim35\%$ with the given $C_{\|}$ and $M_0$ meets the conditions of the existence of the edge states  \cite{Zhou_PRL_2008, Lu_PRB_2010}. However, as seen in Fig.~\ref{1SL_vs_vdW_params}(a), the 1SL film is too far from a
topological phase transition at such $\xi$. With this example we illustrate the strong limitations of the predictive capabilities of the effective continuous model.

Let us now analyze the formation of the SL band structure with the inverted band gap. Starting from well-separated layers, Fig.~\ref{1SL_vs_vdW_bands}(f), and going back to the bulk value $d^{(0)}_{\mathrm{vdW}}$ of the van-der-Waals spacing, Fig.~\ref{1SL_vs_vdW_bands}(a), we retrace the valence bands ($v_1$ with the energy $E_1^v$ and $v_2$ with $E_2^v$) with the predominant contribution coming from the $p_z$ orbitals of Te and the conduction bands ($c_1$ with $E_1^c$ and $c_2$ with $E_2^c$) mainly formed by Bi $p_z$ orbitals, see Supplementary Fig. 2. We derive an 8-band Hamiltonian $H^{\mathrm{1SL}}_{\mathrm{\mathbf{kp}}}$ which is presented in Supplementary Note 3. Its eigenvalues are shown in Fig.~\ref{1SL_vs_vdW_bands} by blue lines, and the corresponding parameters as a function of the $d_{\mathrm{vdW}}$ expansion are depicted in Figures~\ref{1SL_vs_vdW_params}(b) and \ref{1SL_vs_vdW_params}(d), see also Supplementary Fig.~4.

As seen in Fig.~\ref{1SL_vs_vdW_params}(b), at $\bar{\Gamma}$ in the large-$d_{\mathrm{vdW}}$ limitthere are two doubly degenerate energy levels, $E_1^v=E_2^v$ and $E_1^c=E_2^c$. Upon decreasing $d_{\mathrm{vdW}}$, the TLs start to interact primarily by their Te-layer sides to cause the bonding-antibonding splitting of the two degenerate levels: The Te-related energies as a function of $d_{\mathrm{vdW}}$ disperse stronger than those of Bi. Near the bulk value $d^{(0)}_{\mathrm{vdW}}$, the splitting is large enough to invert the order of the $E_2^v$ and $E_1^c$ levels, ensuring the topological phase transition. Thus, the stacking procedure that leads to the 3D TI is based on SL building blocks principally different from the Rashba bilayers used in Ref.~\cite{DasBalatsky}. It is essential that in our case the two Rashba constituents of the block (the stand-alone TLs) bring not only the Rashba-split conduction band but also the valence band, see Supplementary Note~3. Then the gap in the SL (which may be inverted or not) is quite naturally the gap between the valence and conduction bands, in contrast to the scenario of Ref.~\cite{DasBalatsky}, where the band gap in the bilayer block is achieved by a dispersive ``finite quantum tunneling'' between the two Rashba constituents -- the 2D electron gases of the adjacent layers.

Figure~\ref{1SL_vs_vdW_params}(d) shows the behaviour of the inverse effective masses of the chosen bands over the $d_{\mathrm{vdW}}$ interval considered. We find that the conduction-band inverse masses, which are equal in the large-$d_{\mathrm{vdW}}$ limit,  $M_1^c=M_2^c$,  change smoothly with decreasing  $d_{\mathrm{vdW}}$: At  $d^{(0)}_{\mathrm{vdW}}$ the parameter $M_2^c$ becomes twice as large, while $M_1^c$ falls below zero. On the contrary, the valence-band inverse masses ($M_1^v=M_2^v$ in the large-$d_{\mathrm{vdW}}$ limit) ``diverge'' because the band $v_1$ moves down and ``goes through'' the I-orbital dominated bands, and $v_2$ moves up and hybridizes with Te $p_{x,y}$ bands, see Fig.~\ref{1SL_vs_vdW_bands} and  Supplementary Fig. 2. Finally, at $d^{(0)}_{\mathrm{vdW}}$ the parameter $M_2^v$ reaches its large $d_{\mathrm{vdW}}$ limit, while $M_1^v$ becomes negative. Thus, in the topologically non-trivial 1~SL we have $M_2^v\approx C_{\|}-M_{\|}$ and $M_1^c\approx C_{\|}+M_{\|}$, where $C_{\|}$ and $M_{\|}$ are the 1SL parameters of the Hamiltonian~(\ref{Hkp_slab}). At that, the interband coupling of the bands $v_2$ and $c_1$ is equal to $V_{\|}$ of the 4-band \textbf{k$\cdot$p} description. This reveals a close relation between the 4-band and 8-band Hamiltonians. However, already with 8 bands there is an absolute gap (see Fig.~\ref{1SL_vs_vdW_bands}(a)), which is reasonably accurate and quite suitable for the theoretical research on  linear response, Hall conductance, and motion of Dirac fermions in external fields.

\section*{Acknowledgements}
This work was supported by the Spanish Ministry of Economy and Competitiveness MINECO (Project Nos. FIS2013-48286-C2-1-P, FIS2013-48286-C2-2-P, and FIS2016-76617-P), the Basque Country Government, Departamento de Educaci\'{o}n, Universidades e Investigaci\'{o}n (Grant No. IT-756-13) and Saint Petersburg State University (Grant No. 15.61.202.2015).

\section*{Author contributions}
I.A.N. and S.V.E. conceived the idea and designed the research within the projects coordinated by P.M.E. and E.V.C.. S.V.E.
performed atomic structure optimization and GGA band structure calculations. I.A.N. and E.E.K. derived the small-size \textbf{k$\cdot$p} Hamiltonians from \textit{ab initio} wave functions and performed the corresponding LDA-based calculations. I.A.N., S.V.E., and E.E.K. wrote the manuscript. All the authors discussed the results and commented on the manuscript.


\newpage
\clearpage
\newpage
\clearpage
\newpage
\clearpage
\newpage

\end{document}